\documentclass[aps,prc,reprint]{revtex4-1}
\UseRawInputEncoding
\usepackage{epsfig,color}
\usepackage[citecolor=blue,colorlinks=true,linkcolor=blue,urlcolor=blue]{hyperref}
\usepackage{bm}
\usepackage{braket}
\usepackage{amsmath}
\usepackage{tikz,xcolor,hyperref}
\definecolor{lime}{HTML}{A6CE39}
\DeclareRobustCommand{\orcidicon}{%
    \begin{tikzpicture}
    \draw[lime, fill=lime] (0,0)
    circle [radius=0.16]
    node[white] {{\fontfamily{qag}\selectfont \tiny ID}};    \draw[white, fill=white] (-0.0625,0.095)
    circle [radius=0.007];    \end{tikzpicture}
    \hspace{-2mm}}
\foreach \x in {A, ..., Z}{%
    \expandafter\xdef\csname orcid\x\endcsname{\noexpand\href{https://orcid.org/\csname orcidauthor\x\endcsname}{\noexpand\orcidicon}}
    }

\begin{document}
\title{ Intense X-ray laser-induced proton emission from halo nuclei}
\author{Binbing Wu$^{1}$}
\author{Jie Liu$^{1}$}
\email{Corresponding author: jliu@gscaep.ac.cn}
\affiliation{$^{1}$Graduate School, China Academy of Engineering Physics, Beijing 100193, China}
\date{\today}
\begin{abstract}
We investigate the intense X-ray laser-induced proton emission from halo nuclei within the framework of a nonperturbative quantum $S$-matrix approach. We have analytically deduced the angular differential as well as the total multi-photon rates of the proton emissions. For a linearly polarized X-ray laser field, we find that the angular distributions of proton emission sensitively depend on the laser frequency and show an interesting petal structure with increasing the laser frequency as well as the number of absorbed photons. Meanwhile, we find the Coulomb repulsion potential between the proton and the remainder nucleus has a strong hindering effect on the total multi-photon rates of the proton emissions, and leads to the blue shifts of the multi-photon transition frequency.  Moreover, the polarization effects of laser fields on total rates of proton emission have also been addressed. We find that the polarized ellipticity corresponding to the maximum of the total
rates depend on the laser frequency showing a transition from perturbative to nonperturbative proton emission. The underlying mechanism of the above findings is uncovered, and some implications are discussed.
\end{abstract}
\maketitle

\section{Introduction}

Halo nuclei are characterized by a few weakly bound halo nucleons and a more tightly bound core \cite{Hansen,Tanihata,Frederico}. Studies of the halo nuclei have provided important insights into the structure of the nucleus and understanding of nuclear force \cite{Ryberg,Phillips}, and are still a hot research topic so far \cite{Hammer,Hongo,Ayyad,Lopez-Saavedra}. In general, the halo nuclei can be divided into neutron-halos and proton-halos \cite{Frederico}. The first and the most well-studied halo nuclei are the neutron-rich $^{11}{\rm Li}$, which was identified as a two-neutron halo nucleus from the measurement of nuclear cross sections of a high-energy radioactive beam of $^{11}{\rm Li}$ with various target elements \cite{Hansen,Frederico,Tanihata}. The proton-halos nuclei include $^{8}{\rm B}$ \cite{Karataglidis} and $^{26}{\rm P}$ \cite{Loureiro}, presumably the $^{17}{\rm Ne}$ \cite{Kanungo} and $^{100}{\rm Sn}$  \cite{Bielich}, etc.

Photonuclear reactions are an important tool to investigate
nuclei \cite{Zilges}. In 1934, the research field was opened with experiments on the photodisintegration of the deuteron by the $\gamma$ rays from a radioactive source \cite{Chadwick}. Since then, the investigations of the nuclear photoeffect have been systematically carried out \cite{Zilges,Chadwick,Bothe,Rauscher,Mohr,Lan}, which relied on high-energy  $\gamma$ rays from synchrotron radiation, electron bremsstrahlung, or Compton backscattering sources \cite{Assafiri}.
 Some studies suggest that the $\gamma$-ray induced proton emission plays an important role in the synthesis of proton-rich nuclei and in the understanding of stellar nucleosynthesis and galactic chemical evolution \cite{Rauscher,Mohr,Lan}. It is worth noting that the photonuclear reactions induced by the $\gamma$ rays are generally a perturbation process with a one-photon absorption due to the low intensity of the $\gamma$ rays.

 Nowadays, intense X-ray lasers can be generated via high-harmonic generation (HHG) \cite{Gohle,Cingoz,Pupeza,Porat} from intense laser fields or the use of free-electron lasers \cite{McNeil,Allaria,Cavaletto,Ueda}. The rapid advance of intense laser technologies, particularly chirped pulse amplification technique (CPA) \cite{Mourou}, makes laser fields being with a wide range of frequencies, high intensities, and durations \cite{ELi,Guo2018}. For instance, in May 2021, a laser pulse with a peak intensity of about $10^{23}~{\rm W/cm^2}$ was produced \cite{Yoon}. On the other hand, X-ray free-electron lasers already allow to generate coherent
light of up 20 keV in frequency, thereby covering some of the low-energy nuclear excited states \cite{Ueda}. These intense light fields provide an alternative scheme to manipulate the nuclear process \cite{Li2021,Cheng,Wense1,Wense2, Wu2021,Wang2022,Delion2017,Qi2019,Palffy,Ghinescu,Cheng,Qi2020,Friedemann2019,Lv2019,
Wang2020,Liu2021,Kohlfurst,Lv2022,Bekx,Wu2022}, in which the nonperturbative effects of intense light-matter interaction are expected to be essential. For example, theoretical treatments have been made on the possibility of using intense light to excite isomeric $^{229}{\rm Th}$ \cite{Wense1,Wense2,Wu2021,Wang2022}, and to influence $\alpha$ decay \cite{Delion2017,Qi2019,Palffy,Ghinescu}, proton radioactivity \cite{Cheng}, nuclear fission \cite{Qi2020}, or deuteron-triton fusion processes \cite{Friedemann2019,Lv2019,Wang2020,Liu2021,Kohlfurst,Lv2022,Bekx,Wu2022}.
Moreover, these intense light fields may provide a new experimental platform for studies of atomic physics and nuclear physics on the
femto-to nanometer scale \cite{Fu2021}.

A nonperturbative quantum $S$-matrix approach provided a state-of-the-art and convenient method to address intense light-matter interaction \cite{Reiss,Joachain}.
 This approach has been extremely successful in describing highly nonperturbative process of atoms to intense laser fields such as
multiphoton ionization and high harmonic generation \cite{Reiss,Joachain}. The main idea of the quantum $S$-matrix approach is the nonperturbative treatment on the Dyson expansion and to approximate the continuum states by Volkov states \cite{Joachain}. Recently, based on this approach, the two-photon and three-photon absorption processes of deuteron induced by the intense $\gamma$ rays has been theoretically calculated \cite{Li2021}. Moreover, proton emission from halo nuclei via laser-assisted nuclear photoeffect (i.e., the combined a $\gamma$ ray and low-frequency intense laser) has been  investigated \cite{Anis,Anis2,Peter}. These studies have shown that rich sideband structures arise in the photo-proton energy spectra,
and the total cross section of the photo-proton emission is not almost affected by the presence of the laser field \cite{Anis,Anis2,Peter}.

In this present work, inspired by the rapid advance of intense X-ray laser, we investigate proton emission from
halo nuclei induced by intense X-ray laser with arbitrary polarization. Regarding the nuclear species, we focus on one-proton halo nuclei because they have low proton
binding energies \cite{Minamisono,Fukuda,Smedberg}. Symbolically, the process under investigation may be written as
\begin{eqnarray}
^A_Z[A{\rm p}]+n \omega\longrightarrow ^{A-1}_{Z-1}A+^1_1{\rm p},
\end{eqnarray}
where $n$ denotes the number of laser photons involved. The nonperturbative quantum $S$-matrix approach
is exploited to calculate the emission rate of the proton.

The work is organized as follows.  In Sec. II, we present our theoretical framework. In Sec. III, we present
our main results and discussions.  Sec. IV is the main conclusions and outlooks.

\section{Theoretical framework}

In this section, we outline a theoretical model to describe the proton emission from halo nuclei in the presence of a background laser field. The laser field is assumed to be an arbitrarily polarized, monochromatic
wave of frequency $\omega$. It is described by the uniform and time varying vector potential
\begin{eqnarray}
\bm{A}(t)=\frac{A_0}{\sqrt{1+\delta^2}}\left[{\rm cos}(\omega t)\vec{e}_z+\delta {\rm sin}(\omega t)\vec{e}_y\right],
\label{Vector optential}
\end{eqnarray}
where $A_0$ is the maximal amplitude and $0\leq\delta\leq1$ represents the polarized ellipticity \cite{Li2015}.
Note that in Eq. (\ref{Vector optential}) the dipole approximation \cite{Joachain} has been applied, which disregards the spatial field dependence. This approximation is well justified because, for the laser parameters under consideration, the scale of the spatial field variations (i.e., the laser wavelength $\lambda=2\pi c/\omega$, where $c$ is the vacuum light speed) is much larger than both
nuclear size (fm) and laser-driven quiver amplitude of nuclei [$r_e=q A_0/(m\omega)$, where $q$ and $m$ are the charge and rest mass of the nucleus, respectively].

\subsection{Derivation of the transition amplitude~$T^{\rm SFA}_{\bf p}$}
Under the dipole approximation, a two-body total Hamiltonian of one-proton halo nuclei in a laser field can be separated into a center-of-mass part and a relative-motion part \cite{Lv2019,Liu2021,Wu2022}. Since it is the proton emission from halo nuclei that is of interest, the latter part would be focused on and reads
\begin{eqnarray}
H=\frac{\hat{\bm{p}}^2}{2\mu}+V_{\rm N}(r)+V_{\rm C}(r)+\hat {V}_{\rm L}(t),
\end{eqnarray}
where $\hat{\bm{p}}$ is the momentum operator of the relative motion, ${\mu}=m_{\rm p}M/(m_p+M)$ ($m_{\rm p}$ and $M$ are the rest mass of the proton and the remainder nucleus, respectively) is the reduced mass of one-proton halo nuclei, $V_{\rm N}$ is the phenomenological nuclear potential between proton and remainder nucleus, $V_{\rm C}(r)$ is
Coulomb repulsive potential between the proton and the remainder nucleus, and
$V_{\rm L}$ is the interaction potential between the halo nuclei and the laser fields.
In the Coulomb gauge, the interaction potential $V_{\rm L}$ is given as
\begin{eqnarray}
\hat {V}_{\rm L}(t)=-\frac{q_{\rm eff}}{\mu}\bm{A}(t)\cdot\hat{\bm{p}}+\frac{q_{\rm eff}^2\bm{A}^2(t)}{2\mu},
\end{eqnarray}
where $q_{\rm eff}=(q_{\rm p}M-q_{\rm M}m_{\rm p})/(M+m_{\rm p})$ is an effective charge \cite{Lv2019}. Here $q_{\rm p}$ and $q_{\rm M}$ are charges of the proton and the remainder nucleus, respectively.

If at time $t_0$ the halo proton is in its ground state $\ket{\psi_i}$, then the probability amplitude of finding the proton later in a final plane-wave state $\ket{\psi_{\bm p}}$ is
\begin{eqnarray}
T_{\bm p}(t,t_0)=\bra{\psi_{\bm p}}U(t,t_0)\ket{\psi_i},
\label{M_matrix}
\end{eqnarray}
with $\psi_{\bm p}(r)=(2\pi\hbar)^{-3/2}{\rm e}^{{\rm i}\bm{p}\cdot\bm{r}/\hbar}$, where ${\bm p}=(p_x, p_y , p_z)$ is the final momentum of a proton. $U(t,t_0)$ is
the time-evolution operator corresponding to the total Hamiltonian $H$. Note that because of the complexity of nuclear potential $V_{\rm N}$, the analytical form for $U(t,t_0)$ is not simple. For convenience, let us also define $H_0=\hat{\bm{p}}^2/2\mu +V_{\rm N}+V_{\rm C}(r)$ and $H_L=\hat{\bm{p}}^2/2\mu +V_{\rm C}(r)+\hat{V}_{\rm L}(t)$ both being part of the total relative motion Hamiltonian $H$, respectively. However, the time-evolution
operators $U_0$ and $U_{\rm L}$corresponding to the Hamiltonian $H_0$ and $H_L$ are much simpler, respectively.

Using the Dyson expansion, the time-evolution
operators $U_0$ and $U_{\rm L}$ satisfy the following equation \cite{Reiss,Joachain}
\begin{eqnarray}
U(t,t_0)=U_0(t,t_0)-\frac{{\rm i}}{\hbar}\int^t_{t_0}U(t,t_1)\hat{V}_{\rm L}(t_1)U_0(t_1,t_0){\rm d}t_1.\nonumber\\
\label{U_expansion}
\end{eqnarray}
By inserting Eq. (\ref{U_expansion}) into Eq. (\ref{M_matrix}), the probability amplitude $T_{\bm p}$ may be written as
\begin{eqnarray}
T_{\bm p}(t,t_0)=-\frac{{\rm i}}{\hbar}\int^t_{t_0}\bra{\psi_{\bm p}}U(t,t_1)\hat{V}_{\rm L}(t_1)U_0(t_1,t_0)\ket{\psi_i}{\rm d}t_1.\nonumber\\
\label{M_matrix_U}
\end{eqnarray}
Note that the $\bra{\psi_{\bm p}}U_0(t,t_0)\ket{\psi_i}=0$ due to the orthogonality of the discrete states and the plane-wave states. In the spirit of strong field approximation (SFA) \cite{Reiss,Joachain}, the total time-evolution operator $U(t,t_1)$ inside the Eq. (\ref{M_matrix_U}) is approximated by $U_{\rm L}(t,t_1)$, then the transition amplitude $T_{\bm p}$  is further written as
\begin{eqnarray}
T_{\bm p}(t,t_0)&\approx & T_{\bm p}^{\rm SFA}(t,t_0)\nonumber\\
&=&-\frac{{\rm i}}{\hbar}\int^t_{t_0}\bra{\psi_{\bm p}}U_{\rm L}(t,t_1)\hat{V}_{\rm L}(t_1)U_0(t_1,t_0)\ket{\psi_i}{\rm d}t_1\nonumber\\
&=&-\frac{{\rm i}}{\hbar}\int^t_{t_0}\bra{\psi_{\bm p}^{\rm CV}(t_1)}\hat{V}_{\rm L}(t_1)\ket{\psi_i(t_1)}{\rm d}t_1.\nonumber\\
\label{M_SFA}
\end{eqnarray}
This approximation that neglects the nuclear potential in the continuum state is justified for laser-induced proton emission because the nuclear potential $V_{\rm N}(r)$ is of short range.

Using the Coulomb-Volkov solution \cite{Jain,Rosenberg} of a particle, the $\psi_{\bm p}^{\rm CV}(r,t_1)$ in the velocity gauge can be read as
\begin{eqnarray}
\psi_{\bm p}^{\rm CV}(r,t_1)=\frac{{\rm e}^{{\rm i}\bm{p}\cdot\bm{r}/\hbar}\chi({\bm k,r})}{(2\pi\hbar)^{3/2}}
{\rm exp}\left[-\frac{\rm i}{2\mu \hbar}\int_{-\infty}^{t_1}\left[\bm{p}-q_{\rm eff}\bm{A}_{\rm L}(t)\right]^2 d t\right]\nonumber\\
\label{psiCV}
\end{eqnarray}
with
\begin{eqnarray}
\chi({\bm k,r})={\rm e}^{-\pi \eta(k)/2}\Gamma(1+{\rm i}\eta(k)) _1{\rm F}_1\left[-{\rm i}\eta(k),1;i(kr-{\bm k}\cdot{\bm r})\right],\nonumber
\end{eqnarray}
where ${\bm k}={\bm p}/\hbar$ is the wave number of a particle, $\eta(k)=q_{\rm p}q_{\rm M}\mu/(4\pi \epsilon_0
\hbar^2 k)$ is the Sommerfeld parameter, $_1{\rm F}_1$ is the confluent hypergeometric function, and $\Gamma$ is the Gamma function \cite{Abramowitz}.

The initial state $\psi_i$ is an eigenstate of field-free Hamiltonian $H_0$ and can be approximated as a stationary state,
\begin{eqnarray}
\psi_i(r,t_1)\approx\phi_i(r){\rm e}^{{\rm i}E_{\rm b}t_1/\hbar}
\label{psii}
\end{eqnarray}
with the nuclear binding energy $-E_{\rm b}$ \cite{Anis,Anis2,Peter}.
Assuming that the halo proton is in an $s$ state, the space-dependent part can be expressed approximately in Yukawa form \cite{Fukuda} as
\begin{eqnarray}
\phi_i(r)=\frac{c_{0}}{\sqrt{4\pi}}\frac{{\rm e}^{-\beta r}}{\beta r}
\label{phi0}
\end{eqnarray}
with $\beta=1/(\sqrt{2}R_{\rm rms})$ and $c_0=\sqrt{2}\beta^{3/2}$, where $R_{\rm rms}$ is the rms-radius of the proton halo.
By inserting the Eqs. (\ref{psiCV},\ref{psii},\ref{phi0}) into Eq. (\ref{M_SFA}), one can find that
\begin{eqnarray}
 T_{\bm p}^{\rm SFA}(t,t_0)&=&-\frac{{\rm i}}{\hbar}\int^t_{t_0}{\rm d}t_1 \int {\rm d}^3 r  {\psi_{\bm p}^{\rm CV}}^*(r, t_1)\hat{V}_{\rm L}(t_1)\nonumber\\
 &\times&\phi_i(r){\rm e}^{{\rm i}E_{\rm b}t_1/\hbar}.\nonumber\\
 \label{Mp_SFA}
\end{eqnarray}

\subsection{Analytical evaluation of the scattering amplitude $S^{\rm SFA}_{\bf p}$ matrix}

The relative coordinate $r$ of two particles in a laser field is much smaller than the radius $R$ of a nucleon (i.e., $r\ll R$) due to the large mass of nucleon, the approximation is justified:
$\chi({\bm k,r})\approx \chi({\bm k,r})|_{r=0}=\chi_{\rm C}( k)=\sqrt{\frac{2\eta(k)}{{\rm exp}[2\pi\eta(k)]-1}}$,
which is the square root of the so-called Coulomb factor (CF) \cite{Peter}. Then, the final state of a particle can be approximated as
\begin{eqnarray}
\psi_{\bm p}^{\rm CV}(r,t_1)&\approx&\psi_{\bm p}^{\rm CVA}(t_1)=\frac{{\rm e}^{{\rm i}\bm{p}\cdot\bm{r}/\hbar}\chi_{\rm C}( k)}{(2\pi\hbar)^{3/2}}\nonumber\\
&\times&{\rm exp}\left[-\frac{\rm i}{2\mu \hbar}\int_{-\infty}^{t_1}\left[\bm{p}-q_{\rm eff}\bm{A}_{\rm L}(t)\right]^2 d t\right]\nonumber.\\
\label{psiCVA}
\end{eqnarray}
 Putting Eq. (\ref{psiCVA}) in Eq. (\ref{Mp_SFA}) and using partial integration, one can find
 \begin{eqnarray}
 T_{\bm p}^{\rm SFA}(t,t_0)&\approx&\frac{{\rm i}}{(2\pi\hbar)^{3/2}\hbar}\tilde{\phi}_i(\bm p)\chi_{\rm C}( k)\left(\frac{{\bm p}^2}{2\mu}+E_{\rm b}\right)\nonumber\\
 &\times&\int^t_{t_0}{\rm d}t_1 {\rm e}^{\frac{{\rm i}}{\hbar}\left[\int_{-\infty}^{t_1}{\rm d}\tau V_{\rm L}(\tau)
 +\left(\frac{{\bm p}^2}{2\mu}+E_{\rm b}\right)t_1\right]},\nonumber\\
 \label{Mp_SFAA}
\end{eqnarray}
where $\tilde{\phi}_i({\bm p})=\int d^3 r\phi_i(r){\rm e}^{{\rm i}\bm{p}\cdot\bm{r}/\hbar} $ is the Fourier transform of the bound halo state.
The integration over time in the exponent inside Eq. ({\ref{Mp_SFAA}) can be calculated, see appendix \ref{Ap} for details.

 The transition amplitude can then be written in the following form in the longer-time limit:
\begin{eqnarray}
 S_{\bm p}^{\rm SFA}&\equiv& \lim_{t_0\to-\infty \atop t\to+\infty}T_{\bm p}^{\rm SFA}(t,t_0)\nonumber\\
 &\approx&\frac{{2\pi\rm i}}{(2\pi\hbar)^{3/2}}\tilde{\phi}_i(\bm p)\chi_{\rm C}( k)\left(\frac{{\bm p}^2}{2\mu}+E_{\rm b}\right)\nonumber\\
 &\times&\sum^{+\infty}_{n=n_0}\tilde{J}_n(\alpha,-\beta,\eta_0)\delta\left(\frac{{\bm p}^2}{2\mu}+E_{\rm b}+U_{\rm p}-n\hbar\omega\right),\nonumber\\
\end{eqnarray}
where $n_0$ is the smallest integer that satisfies the energy conservation condition, $U_{\rm p}$ is the ponderomotive energy \cite{Joachain}, and $\tilde{J}_n$ is the generalized Bessel function (see appendix \ref{Ap} for details). For each laser photon number $n$, because of the $\delta$ function, the absolute value of the emitted particle momentum be read as
\begin{eqnarray}
 p_n=\sqrt{2\mu(n\hbar\omega-U_p-E_b)}.
\end{eqnarray}

\subsection{The proton emission rates}

\begin{figure*}[!t]
\centering
\includegraphics[width=\linewidth]{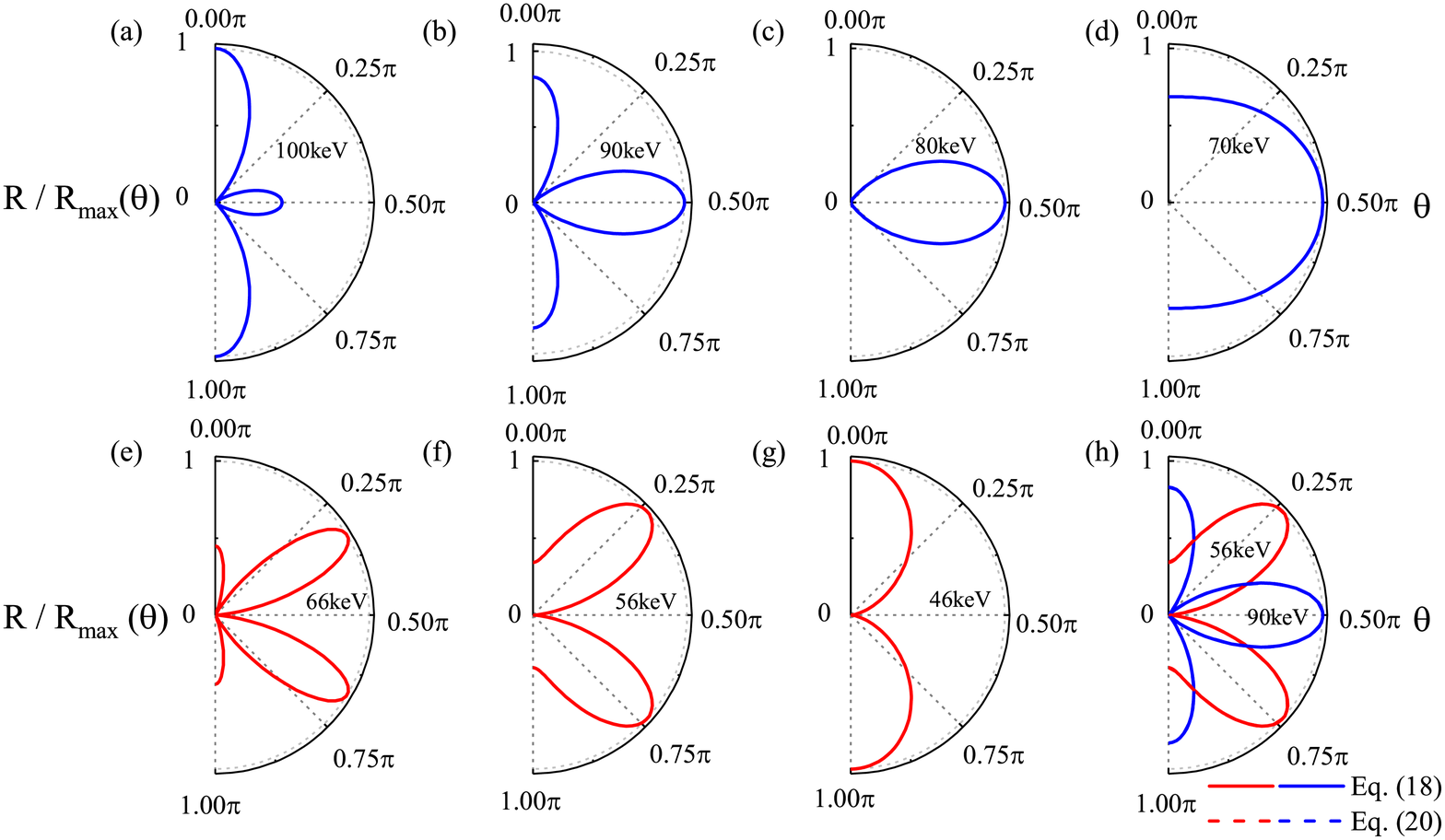}
\caption{(Color online) Angular distributions of the proton emission from halo nuclei $^8_5{\rm B}$ with varied field frequencies $\omega$ for the fixed field intensity $I=10^{23}~{\rm W/cm^2}$. (a)-(d) (top row) represent the process of two-photon absorption while (e)-(g) (bottom row) represent three-photon absorption since the threshold energy of proton emission~$E_{\rm b}=137~{\rm keV}$. (h) is the comparison of the results calculated by Eqs. (\ref{d_n_rate}) (solid lines) and (\ref{app_R}) (dashed lines), respectively. Note that the solid lines are almost identical to the dashed lines in (h). The angle $\theta$ in each figure is the polar angle with respect to the $+z$ direction. Each distribution is normalized to its own peak value. }
\label{figure1}
\end{figure*}

The differential rates of laser-induced proton emission can be obtained by integrating out the final $p$ in each solid angle
\begin{eqnarray}
 \frac{dR}{d\Omega}=\int_0^{+\infty}\frac{ p^2|S_{\bm p}^{\rm SFA}|^2}{T}{\rm d}p\equiv\sum^{+\infty}_{n=n_0}\frac{dR_n}{d\Omega},
\label{d_rate}
\end{eqnarray}
with the differential rate of the $n$-photon absorption
\begin{eqnarray}
 \frac{dR_n}{d\Omega}&=&\frac{2\pi\mu}{(2\pi\hbar)^3}p_n\tilde{\phi}_i^2(\bm p_n)\chi_{\rm C}^2(k_n)(n\hbar\omega-U_{\rm p})^2\nonumber\\
&\times&|\tilde{J}_n(\alpha_n,-\beta,\eta_{0n})|^2,
\label{d_n_rate}
\end{eqnarray}
where $T$ is the period of the laser field.
Then, the total proton emission rate can be obtained by integrating the differential rates over all solid angles:
\begin{eqnarray}
 R=\int_0^{2\pi}d\varphi \int_0^{\pi}d\theta \frac{dR}{d\Omega}{\rm sin}\theta\equiv\sum^{+\infty}_{n=n_0}R_n,
\label{rate}
\end{eqnarray}
where $R_n$ presents the partial emission rate of the $n$-photon absorption.

\section{Numerical results and discussions}
For the initial wave function $\psi_i$ of the halo proton, there are two parameters $E_{\rm b}$ and $R_{\rm rms}$.
We would consider throughout the one-proton halo isotope $^8_5{\rm B}$ as an example due to the very low proton separation energy of $E_{\rm b}=137~ {\rm keV}$ \cite{Anis,Anis2,Peter,Minamisono,Fukuda,Smedberg}. The value $R_{\rm rms}\approx4.73~{\rm fm }$ is used as the root-mean-square distance in the $^7_4{\rm Be}$-proton system \cite{Anis,Anis2,Peter,Minamisono,Fukuda,Smedberg}. Moreover, we focus on the case of the vector potential ${\bm A}(t)$ linearly polarized along the $+z$ axis, [ i.e., $\delta=0$ in Eq. (\ref{Vector optential}) ] throughout our work unless explicitly stated otherwise. Based on Eqs. (\ref{d_rate}) and (\ref{rate}), we have performed numerical calculations of laser-induced proton emission from the halo nuclei.

\subsection{Angular distributions of the proton emission}

We plot the angular distributions of the proton emission from halo nuclei $^8_5{\rm B}$ with different field frequencies $\omega$ for field intensity $I=10^{23}~{\rm W/cm^2}$ and the results are shown in Fig. \ref{figure1}. The Figs. \ref{figure1} (a)-(d) represent the two-photon absorption processes while the three-photon ones for the Figs. \ref{figure1} (e)-(f) since the threshold
 energy of the proton emission $E_{\rm b}=137$~keV. The angular distributions show striking laser frequency dependence and rather an interesting petal structures in Fig. \ref{figure1}.

For the two-photon absorption process, the proton emission is mostly along angle $\theta=0$ and $\pi$ with a small lobe along $\theta=\pi/2$ for $\omega=100~{\rm keV}$ in Fig. \ref{figure1} (a). For $\omega=80~{\rm keV}$ in Fig. \ref{figure1} (c), however, the proton emission is mostly  perpendicular to the field polarization (i.e., $\theta=\pi/2$). For the three-photon absorption process, differently, the peaks of distributions are  neither perpendicular nor parallel to the field polarization without the small lobe along $\theta=\pi/2$ in Figs. \ref{figure1} (e)-(f). It is also interesting that if the frequency slightly decreases from $\omega={\rm 70~keV}$ to 66~keV, the distributions show a significant difference in Figs. \ref{figure1} (d)-(e), which corresponds to the transition from two-photon absorption to three-photon absorption. Note that the similar angular distributions of the proton emission from the deuteron are also confirmed in theory, which the threshold energy of proton is 2.22 MeV \cite{Li2021}. The angular distributions of the proton from halo nuclei $^8_5{\rm B}$ might have more implications in ultra-intense laser facilities compared to
 proton emission from a deuteron.

  In order to further understand the above results, the approximate expansion of the Bessel function is exploited in Eq. (\ref{d_rate}) for the condition $\beta\ll1$ \cite{Li2021,Abramowitz}.
\begin{figure}[!t]
\centering
\includegraphics[width=\linewidth]{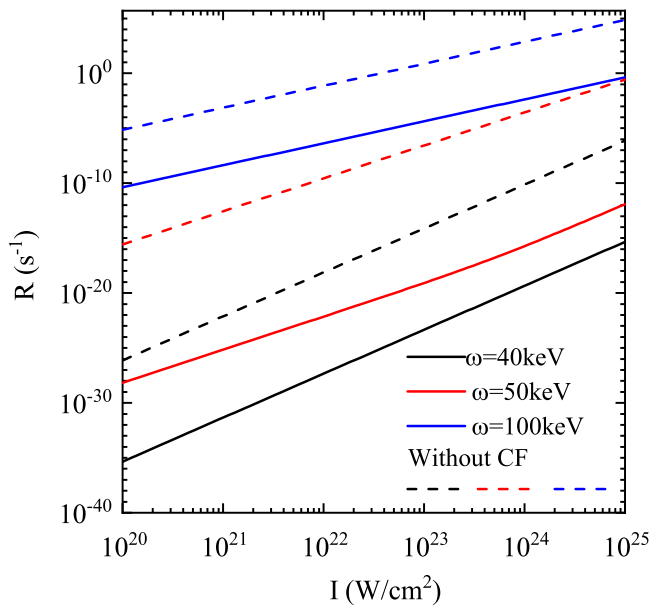}
\caption{(Color online) Total rates of proton emission as a function of laser intensity for different field frequencies: results with (solid lines) and without (dashed lines) CF. Note that the laser field is linearly polarized. }
\label{figure2}
\end{figure}
Then,
the differential rates for two and three-photon absorption can be approximated as
\begin{eqnarray}
\frac{dR_2}{d\Omega}&\propto& 1-8\left(2-\frac{E_{\rm b}}{\omega}\right){\rm cos^2}\theta+16\left(2-\frac{E_{\rm b}}{\omega}\right)^2{\rm cos^4}\theta\nonumber,\\
\frac{dR_3}{d\Omega}&\propto&{\rm cos^2}\theta-\frac{8}{3}\left(3-\frac{E_{\rm b}}{\omega}\right){\rm cos^4}\theta+\frac{16}{9}\left(3-\frac{E_{\rm b}}{\omega}\right)^2{\rm cos^6}\theta\nonumber,\\
\label{app_R}
\end{eqnarray}
respectively. Eq. ({\ref{app_R}}) clearly shows that angular distributions are only sensitively dependent on the laser frequency. Fig. \ref{figure1} (h) displays the comparison of
approximate results (dash lines) calculated by  Eq. ({\ref{app_R}}) and exact ones (solid lines) with Eq. (\ref{d_rate}). The blue and red lines represent the two-photon absorption and the three-photon one, respectively. As is shown in Fig. \ref{figure1} (h), the approximate results are almost identical to the exact ones. These results suggest that the approximate expansion [i.e., Eq. ({\ref{app_R}})] is valid under our consideration of laser parameters.

\subsection{Total rates of proton emission with varied laser field parameters: Coulomb effects}

Total rates of proton emission as a function of laser intensity with (solid lines) and without (dashed lines) CF  are shown in Fig. \ref{figure2}. For two fixed frequencies $\omega=40~ {\rm keV}$ and $100~ {\rm keV}$, the total rate increases almost linearly with increasing laser intensity in logarithmic coordinates. As would be expected, the quantitative relationship between total rate, laser intensity, and the number of absorbed photons can be written approximately as $R\propto I^n$ using the Fermi's golden rule in perturbative regime \cite{Joachain,Orear}.
\begin{figure}[!tb]
\centering
\includegraphics[width=\linewidth]{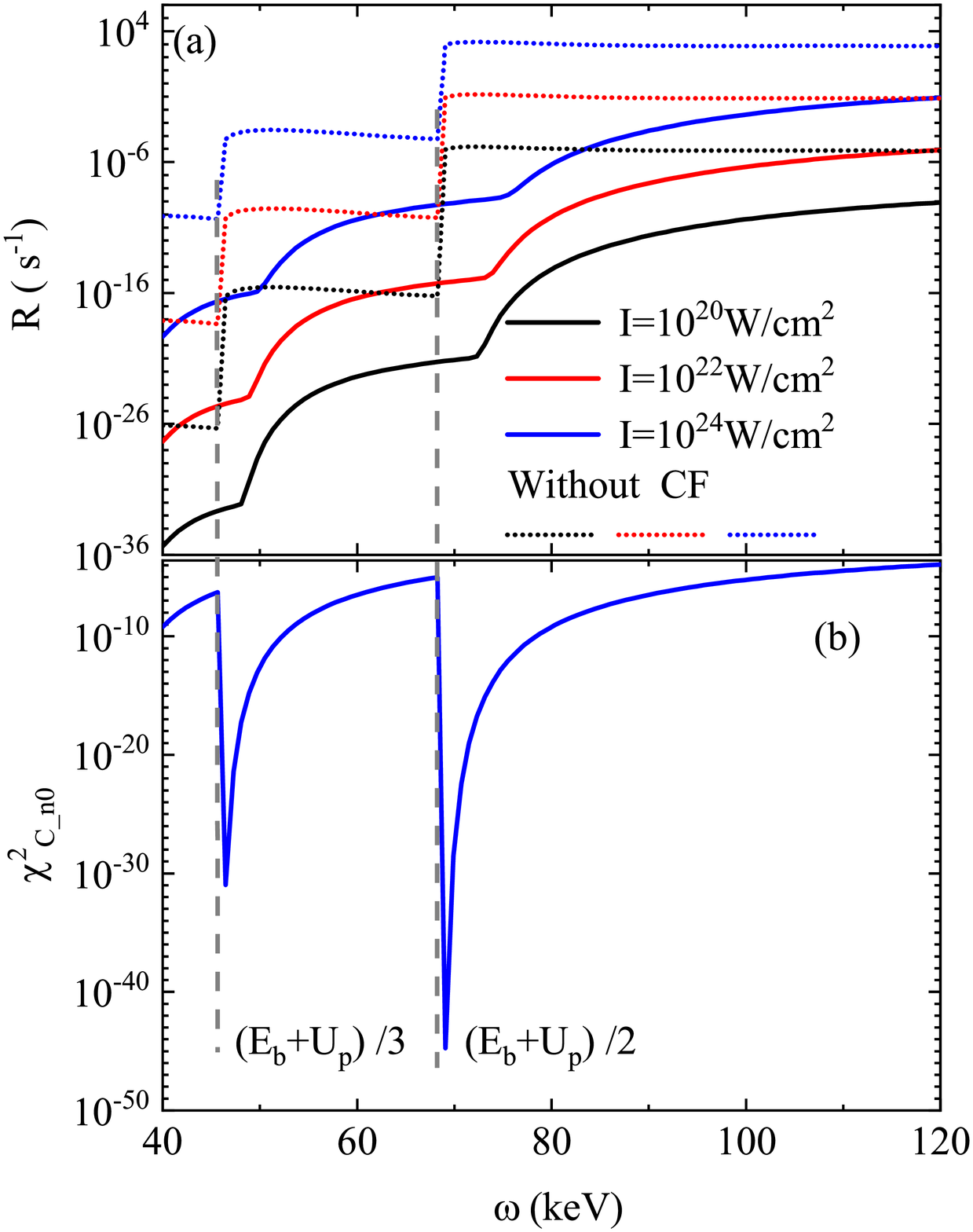}
\caption{(Color online) (a) Total rates of proton emission as a function of laser frequency for different field intensities: results with (solid lines) and without (dashed lines) CF. (b) The corresponding CF  for the smallest number $n_0$ of absorbed photons. Note that the laser field is linearly polarized. The two vertical dotted lines represent $\omega=(E_{\rm b}+U_{\rm p})/3$ and $(E_{\rm b}+U_{\rm p})/2$, respectively.
}
\label{figure3}
\end{figure}
The total rate for the frequency $\omega=50~ {\rm keV}$ in Fig. \ref{figure2}, however, violates  the power law (i.e., $R\propto I^n$) for the dependence on laser intensity. This result implies a nonperturbative signature and further discussions on it will be given later with details in the next subsection.

Moreover, as one can see from solid lines and dash lines in Fig. \ref{figure2}, the CF significantly suppresses the total rate while change slightly the slope of the line for each frequency partly due to the frequency dependence of CF.
These results suggest that Coulomb repulsion potential between the proton and remaining nucleus has a strong hindering effect on the proton emission near the threshold.

We also display the total rates as a function of the laser frequency with (solid lines) and without (dashed lines) CF for different field intensities in Fig. \ref{figure3} (a). We find that the Coulomb repulsion potential has a strong
hindering effect and leads to the blue shifts
of the multi-photon transition frequency in Fig. \ref{figure3} (a). For the case of CF, the total rate increases monotonically with some multi-photon transition points as the laser frequency increases in Fig. \ref{figure3} (a). On the other hand, the total rates exhibits a clear step-like increase for the case without CF.  We plot two vertical dotted lines which represent $\omega=(E_{\rm b}+U_{\rm p})/3$ and $(E_{\rm b}+U_{\rm p})/2$, respectively and find that the vertical line almost coincides with the multi-photon transition frequency of the dash line  but is away from those of the solid line showing a blue shift in Fig. \ref{figure3} (a).

To further understand the blue shift of the multi-photon transition frequency in Fig. \ref{figure3} (a), we plot the corresponding CF for the smallest number $n_0$ of absorbed photons as a function of laser frequency in Fig. \ref{figure3} (b). The CF shows two minimum points which are almost identical to the vertical lines with increasing frequency. For the frequency regime near the minimum points of the curve in Fig. \ref{figure3} (b), the Coulomb repulsion potential shows a hindering effect of local maxima on protons generated by multiphoton. The interplay between the multiphoton process and
the CF effect leads to the results of the solid lines in Fig. \ref{figure3} (a).

It is very interesting to compare the two situations of atomic ionization and proton emission in intense laser field. For the atomic ionization of strong fields, the longe-range Coulomb potential between ions and electrons is usually ignored in the continuum state within the framework of the SFA \cite{Reiss,Joachain}. This approximation for this process is rather justified since the laser field affects electrons in the continuum state far more than Coulomb field \cite{Reiss,Joachain}. In the process of proton emission from nuclei, however, the Coulomb potential more significantly affects proton emission due to the tiny distance (fm) between the proton and remainder nucleus as well as tiny quiver amplitude (fm) of the proton compared to the case of atomic ionization (${\rm \AA}$).
\begin{figure}[!tb]
\centering
\includegraphics[width=\linewidth]{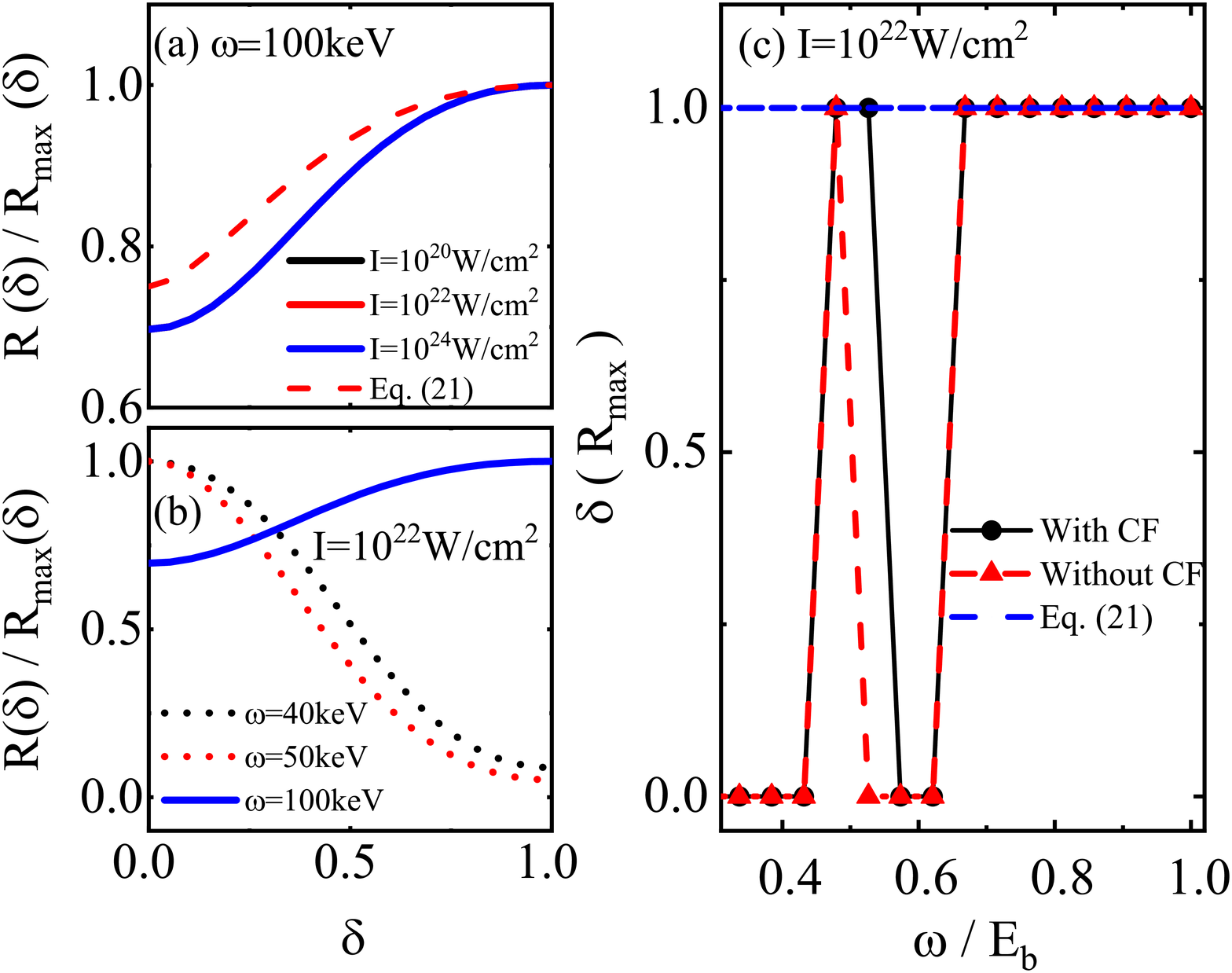}
\caption{(Color online) (a)-(b) Total rates of proton emission as a function of the laser ellipticity $\delta$ for varied intensities and frequencies, respectively. Note that each line is normalized to its maximum value. (c) The laser ellipticity corresponding to the maximum of the total rates as a function of dimensionless variable $\omega/E_{\rm b}$ for the intensity $I=10^{22}{\rm W/cm^2}$: results with (black solid line) and without (red dashed line) CF. The red dashed line in (a) and the blue line in (c) are calculated by Eq. (\ref{ra}), respectively.}
\label{figure4}
\end{figure}

\subsection{The polarization effects of laser fields on the total rates: the transition from perturbative to nonperturbative proton emission}\label{B}

In this section, we focus on the polarization effects [ i.e., $\delta$ in Eq. (\ref{Vector optential}) ] of laser fields on total rates of proton emission. Total rates as a function of the laser ellipticity for different intensities and frequencies are shown in Figs. \ref{figure4} (a) and (b), respectively.

Fig. \ref{figure4} (a) shows that total rates increase monotonically with increasing the laser ellipticity $\delta$ and are insensitive to the laser intensity for the fixed frequency $\omega=100~{\rm keV}$. For the frequencies $\omega=40~{\rm keV}$ and 50 keV, however, the total rates decrease monotonically with increasing the laser ellipticity in Fig. \ref{figure4} (b). The above results suggest a signature of the transition from perturbative to nonpertubative proton emission.

To see clearly the signature of the transition, we also display the laser ellipticity corresponding to the maximum of the total rates as a function of dimensionless variable $\omega/E_{\rm b}$ for the fixed intensity $I=10^{22}{\rm W/cm^2}$: results with (solid lines) and without (dashed lines) CF in Fig. \ref{figure4} (c). One can see from
Fig. \ref{figure4} (c) that the laser ellipticity is 0 or 1 (i.e., linear or circular polarization) for the maximum of the total rates. More specifically, the total rates in low frequency regime (i.e., $\omega/E_{\rm b}<0.45$) reach its maximum at $\delta=0$ and at $\delta=1$ in high frequency regime (i.e., $\omega/E_{\rm b}>0.63$) while at $\delta=0$ or 1 in medium frequency regime.
Moreover, one can also find from Fig. \ref{figure4} (c) that the CF can alter the transition frequency.
\begin{figure}[!tb]
\centering
\includegraphics[width=\linewidth]{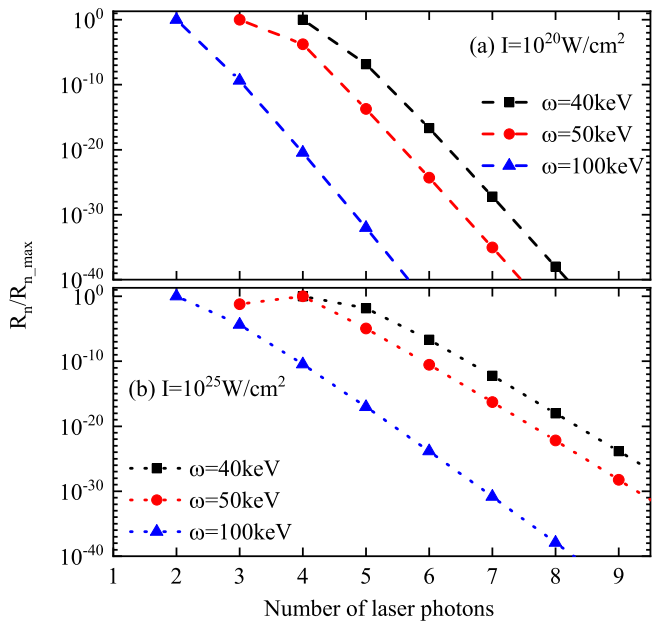}
\caption{(Color online) Distributions of the partial rates $R_n$ of the proton emission with varied numbers $n$ of laser photons for the two fixed intensities (a) $I=10^{20} {\rm W/cm^2}$ and (b) $I=10^{25} {\rm W/cm^2}$, respectively. Each distribution is normalized to its own peak value. Note that the laser field is linearly polarized (i.e., $\delta=0$) and the lines are added just to guide the eye.}
\label{figure5}
\end{figure}

According to the power law for dependence on laser intensity in
perturbation regime \cite{Joachain,Orear}, the total rate for arbitrarily
polarized field [i.e., Eq. (\ref{Vector optential})] would be approximated as
\begin{eqnarray}
&R&\propto\overline{|{\bm E}(t)|}^{2n}\approx\left(\frac{1}{T}\int_0^T |{\bm E}(t)|{\rm d}t\right)^{2E_b/\omega}\nonumber\\
&=&I^{E_b/\omega}
\left[\frac{\left\{{\rm Eli}(1-\delta^2)+\delta {\rm Eli}(1-1/\delta^2)\right\}^2}{\pi^2(1+\delta^2)}\right]^{E_b/\omega},\nonumber\\
\label{ra}
\end{eqnarray}
where the Eli is a function of the complete elliptic integral of the second kind \cite{Abramowitz}. The red dashed line in Fig. \ref{figure4} (a) and the blue line in Fig. \ref{figure4} (c) are calculated by Eq. (\ref{ra}), respectively. One can see from Fig. \ref{figure4} (a) that the red dashed line and the solid line have almost the same trends but quantitatively different for $\omega=100~{\rm keV}$. Meanwhile, Fig. \ref{figure4} (c) shows that the results of the power law [i.e, Eq. (\ref{ra})] are exactly the same as that of nonperturbative $S$-matrix theory in high frequency regime. In low frequency regime, however, the power law fails, which imply a nonperturbative signature. The interplay between the perturbative
and nonperturbative processes might be displayed in medium frequency regime in Fig. \ref{figure4} (c).

These nonperturbative signatures can be further confirmed in Fig. \ref{figure5} where the distributions of the partial rates $R_n$ as a function of the number of laser photons are displayed for different laser field parameters. Note the laser field is linearly polarized (i.e., $\delta=0$) and the lines is added just to guide the eye in Fig. \ref{figure5}. One can see from Fig. \ref{figure5} that for frequency $\omega=50{\rm~keV}$, the scattered points are clearly not in a straight line compared to other frequencies: more interestingly, for intensity
$I=10^{25}{\rm W/cm^2}$, the four-photon rate $R_4$ is greater
than that of the three-photon one $R_3$. Moreover, Fig. \ref{figure5} also shows that the contribution of the partial rates of higher-order photons to total rate increases with increasing intensity.
These results suggest a nonperturbative signature and can not be understood by a simple power law (i.e., $R_n \propto I^n$) in perturbative regime.

\section{Conclusions and outlooks}

In conclusion, the physics of intense X-ray laser-induced proton emission from halo nuclei has been investigated
based on the nonperturbative $S$-matrix theory. We find that the angular distributions of proton emission sensitively depend on the laser frequency and show an interesting petal structure.  Meanwhile, we find the Coulomb repulsion potential between the proton and the remainder nucleus has a strong hindering effect on the total multi-photon rates of the proton emissions, and leads to the blue shifts
of the multi-photon transition frequency. The polarization effects of laser fields of proton emission have also been addressed, in which the signature of the transition from perturbative to nonperturbative process is found and studied thoroughly. Our investigations might have implications for the experiments in ultra-intense laser facilities such as the Extreme Light Infrastructure (ELI) \cite{ELi} and the superintense ultrafast laser facility of Shanghai \cite{Guo2018}.

In this present work, we only focus on the halo nuclei $^{5}{\rm B}$ as an example and extension to other complex halo nuclei like $^{26}{\rm P}$ \cite{Loureiro}, $^{100}{\rm Sn}$ \cite{Bielich}, etc, can be anticipated. On the other hand,
our current results are calculated under the SFA, where the approximation of the Coulomb-Volkov functions is exploited. Discussions beyond the SFA and using precise Coulomb-Volkov functions \cite{Jain,Rosenberg} should be further investigated.
Moreover, we only consider the simple initial state of the halo proton which is a Yukawa form.  Within the framework of halo effective field theory (EFT), a more realistic initial state of the halo proton \cite{Phillips,Hammer} for X-ray laser-induced proton emission is of interest and is a challenging topic worthy of further consideration.

\section*{Acknowledgments}

This work was supported by funding from NSAF No. U1930403.

\appendix

\section{}
\label{Ap}
In this appendix, the temporal integrals in the exponent inside Eq. ({\ref{Mp_SFAA}) can be calculated in detail. In view of temporal integrals in the exponent inside Eq. ({\ref{Mp_SFAA}), by inserting vector potential $\bm{A}(t)$ [i.e.,  Eq. (\ref{Vector optential})] into Eq. ({\ref{Mp_SFAA}), one can obtain
\begin{eqnarray}
 {\rm e}^{\frac{{\rm i}}{\hbar}\left[\int_{-\infty}^{t_1}{\rm d}\tau V_{\rm L}(\tau)
 +\left(\frac{{\bm p}^2}{2\mu}+E_{\rm b}\right)t_1\right]}={\rm e}^{\frac{{\rm i}}{\hbar}
 \left(\frac{{\bm p}^2}{2\mu}+E_{\rm b}+U_{\rm p}\right)t_1}f(t_1)\nonumber\\
 \label{et_SFAA}
\end{eqnarray}
with the time-dependent periodic function
\begin{eqnarray}
 f(t_1)={\rm exp}\left[{\rm i}(\beta {\rm sin}2\omega t_1-\alpha_z {\rm sin}\omega t_1+\alpha_y {\rm cos}\omega t_1)\right].\nonumber\\
\end{eqnarray}
Here, $\beta$, $\alpha_z$, and $\alpha_y$ be defined as
\begin{subequations}
\begin{eqnarray}
\beta=\frac{q_{\rm eff}^2A_0^2(1-\delta^2)}{8\mu\hbar\omega},\\
\alpha_z=\frac{q_{\rm eff}A_0}{\mu\hbar\omega\sqrt{1+\delta^2}}p_z
,\alpha_y=\frac{q_{\rm eff}A_0\delta}{\mu\hbar\omega\sqrt{1+\delta^2}} p_y,
\end{eqnarray}
\end{subequations}
respectively.
Beside, the ponderomotive energy $U_{\rm p}$  is given by
\begin{eqnarray}
 U_{\rm p}=\frac{q_{\rm eff}^2A_0^2(1+\delta^2)}{4\mu},
\end{eqnarray}
which represents the cycle-averaged kinetic energy of the particle in the laser field.
Note that $f(t_1)$ may also be further read as
\begin{eqnarray}
 f(t_1)={\rm exp}\left\{{\rm i}\left[\beta {\rm sin}(2\omega t_1)-\alpha {\rm sin}(\omega t_1-\eta_0)\right]\right\}\nonumber\\
\end{eqnarray}
with $\alpha=\sqrt{\alpha_z^2+\alpha_y^2}$ and $\eta_0={\rm arctan}(\alpha_y/\alpha_z)$.

  By exploiting the Jacobi-Anger identity \cite{Abramowitz}, one can expand $f(t_1)$ into the Fourier series
 \begin{eqnarray}
 f(t_1)=\sum^{+\infty}_{n=-\infty}\tilde{J}_n(\alpha,-\beta,\eta_0){\rm e}^{-{\rm i}n \omega t_1}
 \label{ft_fourier}
\end{eqnarray}
with
\begin{eqnarray}
\tilde{J}_n(\alpha,-\beta,\eta_0)=\sum^{+\infty}_{m=-\infty}J_{n-2m}(\alpha)J_m(-\beta){\rm e}^{{\rm i}(n-2m)\eta_0}\nonumber,\\
\end{eqnarray}
where $J_n$ is a Bessel function of the first kind of integer order $n$ \cite{Abramowitz}.

\end{document}